\documentclass[11pt]{article}\textwidth 6.5in\textheight 9in
\usepackage{amssymb}\usepackage[colorlinks]{hyperref}\usepackage{color}
	\usepackage{stmaryrd}\usepackage{mathrsfs}
	\usepackage{graphicx}
	\usepackage{amsmath}
	
	\usepackage{braket}
\usepackage{caption}

\begin{document}
\setlength{\captionmargin}{27pt}

\newcommand\hreff[1]{\href {http://#1} {\small http://#1}}
\newcommand\trm[1]{{\bf\em #1}} \newcommand\emm[1]{{\ensuremath{#1}}}
\newcommand\prf{\paragraph{Proof.}}\newcommand\qed{\hfill\emm\blacksquare}

\newtheorem{thr}{Theorem} 
\newtheorem{lmm}{Lemma}
\newtheorem{cor}{Corollary}
\newtheorem{con}{Conjecture} 
\newtheorem{prp}{Proposition}

\newcommand\QC{\mathbf{QC}} 
\newcommand\C{\mathbf{C}} 

\newtheorem{blk}{Block}
\newtheorem{dff}{Definition}
\newtheorem{asm}{Assumption}
\newtheorem{rmk}{Remark}
\newtheorem{clm}{Claim}
\newtheorem{exm}{Example}
\newtheorem{exc}{Exercise}

\newcommand\Ks{\mathbf{Ks}} 
\newcommand{\ab}{a\!b}
\newcommand{\yx}{y\!x}
\newcommand{\yux}{y\!\underline{x}}

\newcommand\floor[1]{{\lfloor#1\rfloor}}
\newcommand\ceil[1]{{\lceil#1\rceil}}

\renewcommand\do[1]{\overline{\overline{#1}}}

\newcommand{\bmu}{\boldsymbol{\mu}}

\newcommand{\lea}{<^+}
\newcommand{\gea}{>^+}
\newcommand{\eqa}{=^+}

	\newcommand{\bnu}{\boldsymbol{\nu}}

\newcommand{\lel}{<^{\log}}
\newcommand{\gel}{>^{\log}}
\newcommand{\eql}{=^{\log}}

\newcommand{\F}{\mathbf{F}}
\newcommand{\E}{\mathbf{E}}
\newcommand{\lem}{\stackrel{\ast}{<}}
\newcommand{\gem}{\stackrel{\ast}{>}}
\newcommand{\eqm}{\stackrel{\ast}{=}}

\newcommand\edf{{\,\stackrel{\mbox{\tiny def}}=\,}}
\newcommand\edl{{\,\stackrel{\mbox{\tiny def}}\leq\,}}
\newcommand\then{\Rightarrow}

\newcommand{\kpsi}{\ket{\psi}}
\newcommand{\ktheta}{\ket{\theta}}

\newcommand\Ip{\I_\mathrm{Prob}}
\newcommand\uhr{\upharpoonright}
\newcommand\Hg{\mathbf{Hg}}
\newcommand\Hv{\mathbf{Hv}}
\newcommand\Hbvl{\mathbf{Hbvl}}
\newcommand\Hb{\mathbf{Hb}}
\newcommand\Hoc{\mathbf{Hoc}}
\renewcommand\H{\mathcal{H}}
\newcommand\ml{\underline{\mathbf m}}
\newcommand\mup{\overline{\mathbf m}}
\newcommand\UI{\mathcal{I}}
\newcommand\km{{\mathbf {km}}}\renewcommand\t{{\mathbf {t}}}
\newcommand\KM{{\mathbf {KM}}}\newcommand\m{{\mathbf {m}}}
\newcommand\md{{\mathbf {m}_{\mathbf{d}}}}\newcommand\mT{{\mathbf {m}_{\mathbf{T}}}}
\newcommand\K{{\mathbf K}} \newcommand\I{{\mathbf I}}

\newcommand\II{\hat{\mathbf I}}
\newcommand\Kd{{\mathbf{Kd}}} \newcommand\KT{{\mathbf{KT}}} 
\renewcommand\d{{\mathbf d}} 
\newcommand\D{{\mathbf D}}
\newcommand\Tr{\mathrm{Tr}}
\newcommand\w{{\mathbf w}}
\newcommand\Cs{\mathbf{Cs}} \newcommand\q{{\mathbf q}}
\newcommand\St{{\mathbf S}}
\newcommand\M{{\mathbf M}}\newcommand\Q{{\mathbf Q}}
\newcommand\ch{{\mathcal H}} \renewcommand\l{\tau}
\newcommand\tb{{\mathbf t}} \renewcommand\L{{\mathbf L}}
\newcommand\bb{{\mathbf {bb}}}\newcommand\Km{{\mathbf {Km}}}
\renewcommand\q{{\mathbf q}}\newcommand\J{{\mathbf J}}
\newcommand\z{\mathbf{z}}
\newcommand\Z{\mathbb{Z}}

\newcommand\B{\mathbf{bb}}\newcommand\f{\mathbf{f}}
\newcommand\hd{\mathbf{0'}} \newcommand\T{{\mathbf T}}
\newcommand\R{\mathbb{R}}\renewcommand\Q{\mathbb{Q}}
\newcommand\N{\mathbb{N}}\newcommand\BT{\{0,1\}}
\newcommand\W{\mathbb{W}}
\newcommand\dom{\mathrm{Dom}}
\newcommand\FS{\BT^*}\newcommand\IS{\BT^\infty}
\newcommand\FIS{\BT^{*\infty}}
\renewcommand\S{\mathcal{C}}\newcommand\ST{\mathcal{S}}
\newcommand\UM{\nu_0}\newcommand\EN{\mathcal{W}}

\newcommand{\supp}{\mathrm{Supp}}

\newcommand\lenum{\lbrack\!\lbrack}
\newcommand\renum{\rbrack\!\rbrack}

\newcommand\om{\overline{\mu}}
\newcommand\on{\overline{\nu}}
\newcommand\h{\mathbf{h}}
\renewcommand\qed{\hfill\emm\square}
\renewcommand\i{\mathbf{i}}
\newcommand\p{\mathbf{p}}
\renewcommand\q{\mathbf{q}}
\renewcommand\T{\mathbf{T}}

\title{Two Simple Proofs of M\"{u}ller's Theorem}

\author {Samuel Epstein\\samepst@jptheorygroup.org}

\maketitle

\begin{abstract}
Due to \cite{Muller07,Muller09}, the Kolmogorov complexity of a string was shown to be equal to its quantum Kolmogorov complexity.	Thus there are no benefits to using quantum mechanics to compress classical information. The quantitative amount of information in classical sources is invariant to the physical model used. These consequences make this theorem arguably the most important result in the intersection of algorithmic information theory and physics. The original proof is quite extensive. This paper contains two simple proofs of this theorem. This paper also contains new bounds for quantum Kolmogorov complexity with error.
\end{abstract}

\section{Introduction}

A central topic of investigation in computer science is whether leveraging different physical models can change computability and complexity properties of constructs. In a remarkable result, Shor's factoring algorithm uses quantum mechanics to perform factoring in polynomial time. One question is whether quantum mechanics provides benefits to compressing classical information.
In \cite{Muller07,Muller09}, a negative answer was given, solving open problem 1 in \cite{BerthiaumeVaLa01}.  The (plain) Kolmogorov complexity of a string $x$ is 
 the size of the smallest program to a classical universal Turing machine that can produce $x$. The quantum Kolmogorov complexity of an pure state $\ket{\psi}$, which we call BvL complexity (named after its originators \cite{BerthiaumeVaLa01}), is $\Hbvl(\ket{\psi})$, the size of the smallest mixed quantum state input to a universal quantum Turing machine that produces $\ket{\psi}$ up to arbitrary fidelity. We provide a new simple proof to M\"{u}ller's Theorem. We also present another very simple proof for a slightly less general result using prefix-free Kolmogorov complexity, $\K$. We also introduce new lower bounds on quantum Kolmogorov complexity with error. M\"{u}ller's Theorem is as follows.\\
 
 \noindent\textbf{Theorem}. (\cite{Muller07,Muller09})
 $$
 \C(x)\eqa \Hbvl(\ket{x}\bra{x}).
 $$

\section{Conventions}
We use $\N$, $\Q$, $\R$, $\mathbb{C}$, $\BT$, and $\FS$ to denote natural numbers, rational numbers, reals, complex numbers, bits, and finite strings. Let $X_{\geq 0}$ and $X_{> 0}$ be the sets of non-negative and of positive elements of $X$. When it is clear from the context, we will use natural numbers and other finite objects interchangeably with their binary representations. We use $[A]$ to equal 1 if the mathematical statement $A$ is true and $0$ otherwise.

For positive real functions $f$, by ${\lea}f$, ${\gea}f$, ${\eqa}f$, we denote ${\leq}\,f{+}O(1)$, ${\geq}\,f{-}O(1)$, ${=}\,f{\pm}O(1)$. Furthermore, ${\lem}f$, ${\gem}f$ denotes $< O(1)f$ and $>f/O(1)$. The term and ${\eqm}f$ is used to denote ${\gem}f$ and ${\lem}f$. Plain Kolmogorov complexity is $\C(x)$ and prefix-free Kolmogorov complexity is $\K(x)$. Algorithmic probability is $\mathbf{m}(x)$.

We use the standard model of qubits used throughout quantum information theory. We deal with finite $N$ dimensional Hilbert spaces $\mathcal{H}_N$, with bases $\ket{\alpha_1},\ket{\alpha_2},\dots,\ket{\alpha_n}$. We assume $\mathcal{H}_{n+1}\supseteq\mathcal{H}_n$ and the bases for $\mathcal{H}_n$ are the beginning of that of $\mathcal{H}_{n+1}$. An $n$ qubit space is denoted by $\mathcal{Q}_n = \bigotimes_{i=1}^n\mathcal{Q}_1$, where qubit space $\mathcal{Q}_1$ has bases $\ket{0}$ and $\ket{1}$. For $x\in\Sigma^n$ we use $\ket{x}\in\mathcal{Q}_n$ to denote $\bigotimes_{i=1}^n\ket{x[i]}$. The space $\mathcal{Q}_n$ has $2^n$ dimensions and we identify it with $\mathcal{H}_{2^n}$.
	\begin{dff}[Indeterminate Length Quantum States]
			The separable Hilbert space  $\mathcal{Q} = \bigoplus_{n\in\W}\mathcal{Q}_n$ is the space of indeterminate length quantum states. An example indeterminate length quantum state is
   $$\ket{\psi}= \frac{1}{\sqrt{2}}(\ket{00}+\ket{11011}).
   $$
	\end{dff}
A pure quantum state $\ket{\phi}$ of length $n$ is represented as a unit vector in $\mathcal{Q}_n$. Its corresponding element in the dual space is denoted by $\bra{\phi}$. The tensor product of two vectors is denoted by $\ket{\phi}\otimes\ket{\psi} = \ket{\phi}\ket{\psi} = \ket{\phi\psi}$. The inner product of $\ket{\psi}$ and $\bra{\phi}$ is denoted by $\braket{\psi|\phi}$.

The symbol $\Tr$ denotes the trace operation. The conjugate transpose of a matrix $M$ is denoted by $M^*$.  Projection matrices are Hermitian matrices with eigenvalues in $\{0,1\}$.  For positive semidefinite matrices, $\sigma$ and $\rho$ we say $\sigma \leq\rho$ if $\rho-\sigma$ is positive semidefinite. For positive semidefinite matrices $A$, $B$, $C$, if $A\leq B$ then $\Tr AC\leq \Tr BC$. Mixed states are represented by density matrices, which are, self adjoint, positive semidefinite, operators of trace 1. A semi-density matrix has non-negative trace less than or equal to 1. 

A number is \textit{algebraic} if it is a root of a polynomial with rational coefficients. A pure quantum state $\ket{\phi}$ and (semi)density matrix $\sigma$ are called \textit{elementary} if their real and imaginary components have algebraic coefficients. Elementary objects can be encoded into strings or integers and be the output of halting programs. Therefore one can use the terminology $\K(\ket{\phi})$ and $\K(\sigma)$, and also $\m(\ket{\phi})$ and $\m(\sigma)$. A quantum operation is elementary if its corresponding Kraus operators are elementary.

We say program $q\in\FS$ lower computes positive semidefinite matrix $\sigma$ if, given as input to universal Turing machine $U$, the machine $U$ reads  $\leq\|q\|$ bits and outputs, with or without halting, a sequence of elementary semi-density matrices $\{\sigma_i\}$ such that $\sigma_{i}\leq \sigma_{i+1}$ and $\lim_{i\rightarrow\infty}\sigma_i = \sigma$. A matrix is lower computable if there is a program that lower computes it.

\section{BvL Complexity}
\label{chap:bvl}

	Kolmogorov complexity measures the smallest program to a universal Turing machine that produces a string. Thus it is natural to adapt this notion to defining the complexity of a pure or mixed quantum state $\rho$ to be the shortest program to a universal quantum Turing machine that approximates or produces $\rho$. This definition was introduced  in \cite{BerthiaumeVaLa01} and we call it BvL complexity. 
	
	All quantum Turing machines used in this manuscript are the well formed QTMs defined in \cite{BernsteinVa93}. Well formed QTM preserve length and their time evolution is unitary.
	In this manuscript, BvL complexity is defined with respect to a universal quantum Turing machine introduced in \cite{Muller08}. 
	
	The input, output and auxiliary tapes of $M$ consists of symbols of the type $\Sigma = \{0, 1, \#\}$ . The input is an ensemble $\{p_i\}$ of pure states $\ket{\psi_i}$ of the same length $n$, where $p_i\geq 0$, $\sum_ip_i=1$, and $p_i\in Q_{\geq 0}$. Each pure state $\ket{\psi_i}$ is a complex linear superposition over all inputs of length $n$. Thus the input can be seen as an ensemble of states $\ket{\psi_i\#\#\#\dots}$. This ensemble can be represented as a mixed state $\rho$ of $n$ qubits. The auxiliary tape must contain classical information. The quantum transition function is
	$$\delta: Q\times \Sigma^3\rightarrow \mathbb{C}^{Q\times \Sigma^3\times \{L,R\}^3}.$$
	Note that each complex number must be computable. $Q$ is the set of states, $\Sigma$ is the alphabets on the auxiliary, output, and input tapes, and $\{L,R\}^3$ is the action taken by the three heads. The evolution of $M$ is a computable unitary matrix $u_M$.
	
	There is a start state $\ket{s_C}$ and a final state $\ket{f_C}$. If there exists a $t\in\N$, where during the operation of $M$ input $\rho$,  the control state $M^{t'}_C(\rho)$ is orthogonal to the final state $\ket{f_C}$ for all $t'<t$, with $\bra{f_C}M^{t'}_C(\rho)\ket{f_C}=0$, and $\bra{f_C}M^{t}_C(\rho)\ket{f_C}=1$,   then $M(\rho)$ is defined to be the qubit mixed state $\sigma$ corresponding to the ensemble of pure states determined by ensemble of pure states over the contents of the output tapes at halting time. If one such pure state of the output tape is $\ket{\psi}=\sum_{i=1}^N\alpha_i \ket{s_i \#\#\#\#\#\dots}$, where each $\|s_i\|$ can be different, then the resultant output pure state is $\ket{\tilde{\psi}}=\sum_{i=1}^N\alpha_i\ket{s_i}$. Otherwise, if the the control state evolution is not defined as above, $M(\sigma)$ is undefined.  Thus the output can be a superposition of pure states of different lengths, indeterminate length quantum states. Thus QTMs $M$ can be thought of as partial functions of the following form.
	$$
	M:\bigcup_n\mathcal{Q}_n\rightarrow\mathcal{Q}.
	$$
Thus we only consider \textit{fixed-length} inputs to QTMs $M$. This consists of elements of $\mathcal{Q}$ that are superpositions of basis quantum states $\ket{e_i}$ of the same length.

	  One might argue that this definition with regard to the halting state is too restrictive, but as shown \cite{Muller07},
	for every input $\sigma$ to a QTM that almost halts within a certain computable level of precision, there is another state $\sigma'$ such that $\|\sigma'\|\lea\|\sigma\|$ that makes the universal QTM $\mathfrak{U}$ halt perfectly.
	
	Quantum machines are not expected to produce the target states exactly, only an approximation is required. To measure the closeness of states, the \textit{trace distance} function is used.
	\begin{dff}[Trace Distance and Fidelityt of Quantum States]
		\label{dff:tracedist}
		$D(\sigma,\rho) = \frac{1}{2}\|\sigma-\rho\|_1$, where $\|A\|_1= \Tr\sqrt{A^*A}$.   The trace distance obeys the triangle inequality.
		Fidelity is $F(\sigma,\rho)=\left(\Tr\sqrt{\sqrt{\sigma}\rho\sqrt{\rho}}\right)^2$, with $F(\ket{\psi},\sigma)= \bra{\psi}\sigma\ket{\psi}$ and $1-D(\rho,\ket{\psi})< F(\rho,\ket{\psi})$.
	\end{dff}
		
	\begin{thr}[\cite{Muller08}]
		\label{thr:uqtm}
		There is quantum Turing machine $\mathfrak{U}$ such that for every QTM $M$ and mixed state $\sigma$ for which $M(\sigma)$ is defined, there is mixed state $\sigma'$ such that
		$$D\left(\mathfrak{U}(\sigma'),M(\sigma)\right)< \delta,$$
		for every $\delta\in\Q_{>0}$ where $\|\sigma'\|\lea \|\sigma\|+\K(M,\delta$.
	\end{thr}
	One can define the complexity of a state $\sigma$ with respect to an arbitrary quantum Turing machine.
	\begin{dff}
		The BvL Complexity of mixed state $\rho$  with respect to QTM $M$ and trace distance $\epsilon$ is
		$$\Hbvl_M^\epsilon(\rho) = \min_\sigma \{\|\sigma\| :D(M(\sigma),\rho)< \epsilon\}.$$
		The BvL Complexity of mixed state $\rho$ with respect to QTM $M$ is 
		$$\Hbvl_M(\rho)=\min_\sigma \left\{\|\sigma\|:\forall_k, D(M\left(\sigma,k\right),\rho)<\frac{1}{k}\right\}.$$
	\end{dff}
	Due to Theorem \ref{thr:uqtm} and the fact that the trace distance $D$ follows the triangle inequality, using the universal quantum Turing machine $\mathfrak{U}$, one can define the BvL complexity of a quantum state.  This differs from the original definition in \cite{BerthiaumeVaLa01} where the program must achieve any degree of precision.  
	
	\begin{thr}[\cite{Muller08}]
		For $\delta<\epsilon\in\Q_{>0}$,  universal QTM $\mathfrak{U}$, for every QTM $M$, 
		\begin{itemize}
			\item $\Hbvl_\mathfrak{U}^\epsilon(\sigma) <  	\Hbvl_M^\delta(\sigma) +\K(\epsilon-\delta,M).$
			\item $\Hbvl_\mathfrak{U}(\sigma) < \Hbvl_M (\sigma)+\K(M)$.
		\end{itemize}
	\end{thr}
	\begin{dff}[BvL Complexity] $ $\\
		\label{dff:Hbvl}
		\vspace*{-0.4cm}
		\begin{itemize}
			\item$\Hbvl^\epsilon(\sigma) = \Hbvl_\mathfrak{U}^\epsilon(\sigma) .$
			\item$\Hbvl(\sigma) = \Hbvl_\mathfrak{U}(\sigma)$.
		\end{itemize}
	\end{dff}

 \begin{rmk}
     In fact, $\mathfrak{U}$ is constructed from two different quantum Turing machines. The first machine, $\mathfrak{U}_0$, realizes $\Hbvl^\epsilon$ and the second machine, $\mathfrak{U}_1$ realizes $\Hbvl$. A bit in the input selects which machine to use.
 \end{rmk}
 
\section{An Elementary Approximation of $\mathfrak{U}_0$}

\begin{rmk}
	Let $\mathcal{H}^t_k$ be the linear subspace of $\mathcal{Q}_k$ that spans pure states $\ket{\psi}\in\mathcal{Q}_k$ such that $\mathfrak{U}_0(\ket{\psi})$ is defined and halts in $t$ steps. Due to \cite{Muller07,Muller08}, if $t\neq t'$ then $\mathcal{H}^t_k\perp\mathcal{H}^{t'}_k$.
\end{rmk}
\begin{thr}[\cite{Muller07,Muller08}]
	\label{thr:enumP}
	Given $k,t$, there is an algorithm that can enumerate $\mathcal{H}^t_k$ in the form of elementary projections $\{P_i\}$, such that $\Tr P_iP_j=0$ for $i\neq j$ and $\sum_i P_i$ projects onto $\mathcal{H}^t_k$. Furthermore, all valid inputs $\sigma$ to $\mathfrak{U}_0$ have $\sigma\leq P_i$ for some $P_i$.
\end{thr}

\begin{lmm}
	\label{lmm:psitdk}
	Given $t,k,\delta$ one can compute an elementary quantum operation $\Psi^{t,\delta}_{k}:\mathcal{Q}_k\rightarrow \mathcal{Q}$ such that if $\sigma\in\mathcal{H}^{t}_{k}$ then $D(\Psi^{t,\delta}_{k}(\sigma),\mathfrak{U}_0(\sigma))\leq\delta$.  
\end{lmm}
\begin{prf}
	Let $\Psi=\Psi^{t,\delta}_{k}$. The quantum operation $\Psi$ starts by first  applying quantum operation $\mathcal{E}_1$, which appends $2t$ spaces to the auxiliary, input, and output tape, and then treating the tapes as loops. The start state is appended as well as the header pointer at origin. Then it applies the approximating elementary unitary matrix $\tilde{u}$ corresponding to the unitary matrix $u$ of  $\mathfrak{U}_0$ (with shortened tapes) $t$ times. Then it applies quantum operation $\mathcal{E}_2$, which projects all configurations in the halting state $\ket{q_f}$ of the form $\ket{s_i\#\#\dots}$ to $\ket{s_i}$ and projects configurations with states other than $\ket{q_f}$ to $\lambda\in\mathcal{Q}_0$. So $\Psi(\sigma)=\mathcal{E}_2(\tilde{u}^t\mathcal{E}_1(\sigma)\tilde{u}^{t*})$. It remains to determine the approximation matrix $\tilde{u}$.
	
	Let $\mathcal{C}$ be the finite configuration space. Let $\gamma$ be a parameter to be determined later. First cover $\mathcal{C}$ by elementary mixed states $\rho\in Q$, such that $\max_{\sigma\in\mathcal{C}}\min_{\rho\in Q} D(\sigma,\rho) <\gamma/3$. Next run the algorithm to compute the transition function  of $\mathfrak{U}_0$ long enough to produce unitary matrix $\tilde{u}$ such that for all $\rho\in Q$, $D(u\rho u^*,\tilde{u}\rho\tilde{u}^*) <\gamma/3$. This is possible because the amplitudes of the transition function of $\mathfrak{U}_0$ can be computed to any accuracy. Thus for any $\sigma\in\mathcal{C}$, for proper choice of $\rho\in Q$, by the triangle inequality of trace distance,
	
	\begin{align*}
		D(u\sigma u^t,\tilde{u}\sigma \tilde{u}^*) & < D(u\sigma u^*,u\rho u^*)+D(u\rho u^*,\tilde{u}\rho \tilde{u}^*)+D(\tilde{u}\rho\tilde{u}^*,\tilde{u}\sigma \tilde{u}^*)\\
		& < D(\sigma,\rho)+\gamma/3+D(\rho,\sigma)\\
		& < \gamma.
	\end{align*}
	If $\tilde{u}$ is run twice with any input $\sigma\in\mathcal{C}_n$, the error is bounded by
	\begin{align*}
		D(\tilde{u}^2\sigma \tilde{u}^{2*},u^2\sigma u^{2*}) &< D(\tilde{u}^2\sigma \tilde{u}^{2*},\tilde{u}u\sigma u \tilde{u}) +D(\tilde{u}u\sigma u \tilde{u},u^2\sigma u^{2*})\\
		&< D(u\sigma u^*,\tilde{u}\sigma \tilde{u}^*)+\gamma \\
		&< 2\gamma.
	\end{align*}
	With similar reasoning, one can see that running $\tilde{u}$ a total of $\ell$ times will produce a maximum error of $\gamma\ell$. So $\gamma$ is set to equal $\delta/t$. So for all $\sigma\in\mathcal{Q}_k$,
	\begin{align}
		\label{eq:uutilde}
		D(u^t\mathcal{E}_1(\mathcal{\sigma})u^{t*},\tilde{u}^t\mathcal{E}_1(\mathcal{\sigma})\tilde{u}^{t*})<\delta.
	\end{align}
	If $\sigma\in \mathcal{H}^t_{k,n}$, then $\mathcal{E}_2(u^t\mathcal{E}_1(\mathcal{\sigma})u^{t*})=\mathfrak{U}_0(\sigma)$, so 
	\begin{align*}
		\delta & \geq  D(u^t\mathcal{E}_1(\mathcal{\sigma})u^{t*},\tilde{u}^t\mathcal{E}_1(\mathcal{\sigma})\tilde{u}^{t*})\\
		&\geq D(\mathcal{E}_2(\tilde{u}^t\mathcal{E}_1(\mathcal{\sigma})\tilde{u}^{t*}),\mathcal{E}_2(u^t\mathcal{E}_1(\mathcal{\sigma})u^{t*}))\\
		&=D(\Psi(\mathcal{\sigma}),\mathfrak{U}_0(\sigma)).
	\end{align*}\qed
\end{prf} 
\section{First Proof}
	
	We recall that $\mathcal{Q}$ is the space of indeterminate length quantum states. A semi-density operator $\sigma$ is an
	self adjoint, positive semidefinite, operator over $\mathcal{Q}$ of non negative trace no more than 1. An elementary pure state $\ket{\psi}\in\mathcal{Q}$ is a normalized vector with elementary coefficients residing in a finite number of subspaces $\mathcal{Q}_n$. An elementary semi-density operator can be decomposed into $\sum_{i=1}^Nv_i\ket{\psi_i}\bra{\psi_i}$, where $\ket{\psi_i}$ is an elementary pure state. A semi-density operator $\sigma$ is lower computable if there is an algorithm that outputs a sequence $\{v_i,\ket{\psi}\}_{i=1}^\infty$, where $v_i\in \Q_{\geq 0}$ and $\ket{\psi}$ is elementary and $\sigma = \sum_{i=1}^\infty v_i\ket{\psi_i}\bra{\psi_i}$. The lower complexity of such $\sigma$ is $\ml(\sigma)=\sum\{\m(p):p\textrm{ lower computes }\sigma\}$. There exists a universal lower computable semi-density operator $\mathbf{\bnu}$, such that for all lower computable semi-density operators $\sigma$, $\bnu \gem \ml(\sigma)\sigma$. This is constructed in the standard way in algorithmic information theory.
	\begin{lmm}
		\label{lmm:bnu}
		For $x\in\FS$, $\bra{x}\bnu\ket{x}\eqm \m(x)$.
	\end{lmm}
	\begin{prf}
		Since $\bnu$ is a lower computable semi-density operator, its trace is not more than 1, so $p(x)=\bra{x}\bnu\ket{x}$ is a lower computable semi-measure. So $p(x)\lem \m(x)$. Let $\sigma$ be the lower computable semi density operator $\sigma = \sum_{x\in\FS}\m(x)\ket{x}\bra{x}$. So $\bnu \gem \sigma$ which implies $\bra{x}\bnu\ket{x}\gem \bra{x}\sigma\ket{x} \gem \m(x)$.\qed
	\end{prf}$ $\\
 
 The following lemma is an improvement to results in \cite{Muller09}.
	\begin{lmm}
		\label{lmm:prefixhbvl}
		$\K(x) \lea \Hbvl^\epsilon(\ket{x}\bra{x})+\K(\Hbvl^\epsilon(\ket{x}\bra{x}),\epsilon)-\log (1-1.01\epsilon)$.
	\end{lmm}
	\begin{prf}
		Let $k=\Hbvl^\epsilon(\ket{x}\bra{x})$. We use the algorithm in Theorem \ref{thr:enumP} to enumerate projections $P_i$ for $\mathcal{H}^t_k$, for fixed $k$ and all $t$. We construct the semi-density operator $\nu=2^{-k}\sum_{i}\Psi^{t(i),0.01\epsilon}_k(P_i)$ over the space of indeterminate quantum states $\mathcal{Q}$. 
		
		Let $\sigma$ realize $\Hbvl^\epsilon(\ket{x})$, where $\rho=\mathfrak{U}_0(\sigma)$ in $s$ steps, and $D(\rho,\ket{x})<\epsilon$, and due to Theorem \ref{thr:enumP}, $\sigma\leq P_i$ for some $i$. So due to Lemma \ref{lmm:psitdk}, if $\xi = \Psi^{s,0.01\epsilon}_{k}(\sigma)$, then $D(\xi,\rho)\leq 0.01\epsilon$.  So $D(\xi,\ket{\psi})< 1.01\epsilon$. So, due to the definition of trace distances and fidelity of quantum states, $F(\ket{\psi},\xi )=\bra{\psi}\xi  \ket{\psi}> 1-1.01\epsilon$.
		So, using reasoning analogous to Theorem 9 in \cite{Gacs01}, and due to Lemma \ref{lmm:bnu},
		
			\begin{align*}
			\m(k,\epsilon)\nu  \lem \m(k,\epsilon)2^{-k}\sum_j\Psi^{t(j),0.01\epsilon}_{k}(P_j) &\lem \bnu\\
			\m(k,\epsilon)2^{-k}\Psi^{t(i),0.01\epsilon}_{k}(P_i) &\lem \bnu\\
			\m(k,\epsilon)2^{-k}\Psi^{s,0.01\epsilon}_{k}(\sigma)& \lem \bnu\\
			\m(k,\epsilon)2^{-k}\bra{x}\xi \ket{x} & \lem \bra{x}\bnu\ket{x}\\
			\m(k,\epsilon)2^{-k}(1-1.01\epsilon)& \lem \m(x)\\
			k+\K(k,\epsilon)-\log(1-1.01\epsilon)&\gea \K(x).
		\end{align*}
		
		\qed
		\end{prf}$ $\\

  \begin{prp}
		\label{prp:hbvlehbvl}
		For $k\in\N$, $\Hbvl^{\frac{1}{k}}(\sigma|k)\leq \Hbvl(\sigma)$.
	\end{prp}
	\begin{prf}
        Let $\rho$ be a mixed state such that $\mathfrak{U}_1(\rho|\cdot)$ realizes $\Hbvl(\sigma)$. By Definition \ref{dff:Hbvl}, there is an input $\rho'$ such that $D(\mathfrak{U}_0(\rho'|2k),\mathfrak{U}_1(\rho|2k))<1/2k$ where $\|\rho'\|< \|\rho\|+c_{\mathfrak{U}_1}$. 
        Since $D(\mathfrak{U}_1(\rho|2k),\sigma)<1/2k$, it must be that $D(\mathfrak{U}_0(\rho'|2k),\sigma)<1/k$. So $\Hbvl^{\frac{1}{k}}(\sigma|k)\eqa\Hbvl^{\frac{1}{k}}(\sigma|2k)\leq \Hbvl(\sigma)$. \qed
	\end{prf}
	\begin{prp}
		\label{prp:aKbK}
		For every $c$, there is a $c'$ such that if $a<b+c$ then $a+\K(a)<b+\K(b)+c'$.
	\end{prp}
	\begin{prf}
		So $\K(a-b)< 2\log c+O(1)$. So $\K(a)<\K(b)+2\log c+O(1)$. Assume not, then $b-a+c'< \K(a)-\K(b)+O(1)< 2\log c +O(1)$, which is a contradiction for $c'>2\log  c +O(1)$.\qed
	\end{prf}$ $\\
	
		Lemma \ref{lmm:prefixhbvl} can be used to prove a weaker version of M\"{u}ller's Theorem, as shown in the following corollary.
		\begin{thr}
			$\K(x) \lea \Hbvl(\ket{x}\bra{x})+\K(\Hbvl(\ket{x}\bra{x}))$.
		\end{thr}
		\begin{prf}
			By Lemma \ref{lmm:prefixhbvl},
			$$
			\K(x)\lea \Hbvl^{\frac{1}{2}}(\ket{x}\bra{x})+\K(\Hbvl^{\frac{1}{2}}(\ket{x}\bra{x}),1/2).
			$$
			By Propositions \ref{prp:hbvlehbvl} and \ref{prp:aKbK},
				$$\K(x) \lea \Hbvl(\ket{x}\bra{x})+\K(\Hbvl(\ket{x}\bra{x})).$$
			\qed
		\end{prf}

\section{Second Proof}

	The following new proof of M\"{u}ller's Theorem is self contained, in that the only characterization of the universal QTM $\mathfrak{U}_0$ needed is Theorem \ref{thr:enumP}.
	\begin{thr}[\cite{Muller07,Muller09}]
		$$\C(x)\eqa\Hbvl(\ket{x}\bra{x}).$$
	\end{thr}
\begin{prf}
	$\Hbvl(\ket{x}\bra{x})\lea \C(x)$ because a universal QTM can simulate a classical Turing machine. Let $j=2^{k+5}$ be the precision parameter.  Let $k = \Hbvl^{1/j}(\ket{x}\bra{x}|j)$. By Proposition $k\lea \Hbvl(\ket{x}\bra{x})$. Let $\Psi^{t,\delta}_k(\cdot|j)$ be equal to $\Psi^{t,\delta}(\cdot)$ with the universal QTM $\mathfrak{U}_0$ (and the QTMs it simulates) with $j$ on the auxilliary tape. Using Theorem \ref{thr:enumP}, enumerate all projection operators $P_i$ of $\mathcal{H}^t_k$ (relativized to $j$) for fixed $k$ over all $t$. So $\Tr\sum_i P_i\leq 2^k$. For each $P_i$ enumerated, compute $O_i = \Psi^{t(i),1/j}_k(P_i)$, where each $O_i$ is a positive operator over $\mathcal{Q}$ with $\Tr\sum_i O_i\leq 2^k$.
	
	Assume there is a $k$ qubit input $\sigma\leq P_i$ and a pure state $\ket{\psi}\in\mathcal{Q}_\ell$ such that $D(\mathfrak{U}_0(\sigma,j),\ket{\psi})<1/j$. If $\xi=\Psi^{t(i),1/j}_k(\sigma|j)\leq O_i$ then $D(\xi,\mathfrak{U}(\sigma,j))<1/j$ and by the triangle inequality of trace distances, $D(\xi,\ket{\psi})< 2/j$ and so $1-2/j< F(\xi,\ket{\psi})=\bra{\psi}\xi\ket{\psi}\leq \bra{\psi}O_i\ket{\psi}=\bra{\psi}O^\ell_i\ket{\psi}$, where $O^\ell_i=Q_\ell O_iQ_\ell$, where $Q_\ell$ is the projector onto $\mathcal{Q}_\ell$.
	
	Let $N^\ell_i$ be a projection over $\mathcal{Q}_\ell$ defined from $O^\ell_i$ in the following way. Since $O^\ell_i=\sum_{i}v_i\ket{e_i}\bra{e_i}$ for some orthonormal basis $\{\ket{e_i}\}$, of $\mathcal{Q}_\ell$, we define $N^\ell_i$ to be equal to $\sum_i[1/2\leq v_i]\ket{e_i}\bra{e_i}$. So $\Tr N^\ell_i \leq 2\Tr O^\ell_i\leq 2^{k+1}$. Some simple math shows that if $\bra{\psi}O^\ell_i\ket{\psi}\geq 1-2/j$, then $\bra{\psi}N^\ell_i\ket{\psi}\geq 1 - 4/j=1-2^{-k-3}$. By Lemma \ref{lmm:projao}, there can be only at most $2\Tr N^\ell_i$ classical states $\ket{y}$, $y\in\BT^\ell$, with $\bra{y}N^\ell_i\ket{y}\geq 1  - 2^{-k-3}$. Since $\Tr\sum_{i,j} N^j_i \leq 2^{k+1}$, there only at most $2^{k+1}$ classical strings $\ket{y}$ such that there is a $k$ qubit state $\rho$ such that $D(\mathfrak{U}_0(\rho,j),\ket{y})<j^{-1}$. 
	
	So we define an algorithm that takes in a $k\,{+}\,1$ bit number $b$. For all $i,\ell$,  it enumerates $P_i$, $O_i$, and then each $O^\ell_i$ and  $N^\ell_i$. Then it determines the set $\{\ket{y}\}$ for classical strings $y\in\BT^\ell$ such that $\bra{y}N^\ell_i\ket{y}>1-2^{-k-3}$ for some $i\in \N$. If $\ket{y}$ is the $b$th state discovered with this condition, then return $y$. By the definition of $k$, there is a $k$ qubit input $\rho$ and $P_i\geq\rho$ such that $D(\mathfrak{U}_0(\rho,j),\ket{x})<1/j$, so $x$ will be returned for proper choice of $b$. So $\C(x)\lea \Hbvl(\ket{x})$. \qed	
\end{prf}

	\begin{lmm}
		\label{lmm:projao}
		For a rank $m$ projection matrix $P$ in $\mathbb{C}^n$, assume there is a orthonormal set $\{\ket{e_i}\}_{i=1}^N$ such that $\bra{e_i}P\ket{e_i}>1-1/4m$ for all $i$. Then $N< 2m$.
	\end{lmm}
	\begin{prf}
		Let $Q=I_n-P$. So $\bra{e_i}Q\ket{e_i} \leq 1/4m$.  By the Cauchy Schwarz inequality $|\braket{e_i|Q|e_j}|^2\leq  \bra{e_i}Q\ket{e_i}\bra{e_j}Q\ket{e_j}\leq (1/4m)^2$. So $|\braket{e_i|Q|e_j}|\leq 1/4m$.
		\begin{align*}
			0&=\braket{e_i|e_j}= \braket{e_i|P+Q|e_j}\\
			0&= \braket{e_i|P|e_j}+\braket{e_i|Q|e_j}\\
			|\braket{e_i|P|e_j}| &\leq|\braket{e_i|Q|e_j}|\leq 1/4m.
		\end{align*}
		Let $c_i  = (\braket{e_i|P|e_i})^{1/2}$, where $c_i^2\geq 1-1/4m$. Let $\ket{f_i}=c_i^{-1}P\ket{e_i}$ be a normalized vector. So for $i\neq j$, 
		$$	|\braket{f_i|f_j}|\leq |\braket{e_i|P|e_j}|/(c_ic_j)\leq (1/4m)/(1-1/4m)\leq m^{-1/2}/2. $$	
The following reasoning is due to \cite{Tao24}. Suppose for contradiction $N\geq 2m$. We consider the $2m\times 2m$ Gram matrix $(\braket{f_i|f_j})$, $1\leq i,j\leq 2m$. This matrix is positive semi-definite with rank at most $m$. Thus if one subtracts off the identity matrix, it has an eigenvalue of $-1$ with multiplicity at least $m$. Taking Hilbert-Schmidt norm, we conclude
		$$
		\sum_{1\leq i,j\leq 2n;i\neq j}|\braket{f_i,f_j}|^2 \geq m.
		$$
		But the left-hand side is at most $2m(2m-1)\frac{1}{4m}=m-\frac{1}{2}$, giving the desired contradiction.\qed
	\end{prf}
\section{The Multiverse}
Quantum computers have an interesting interpretation with respect to the Many Worlds Theory. A quantum computer is realized by a number of qubits which can implemented in a number of ways such as trapped ions that behave as magnets. The qubits are isolated from the outside environment to make the decoherence time as long as possible. When the quantum computation begins, unitary transforms are performs on the qubits, which in the context of the Many Worlds Theory, causes an exponential branching on worlds, each containing a different qubit value. The operations of the quantum computer cause interference effects between the branches until a measurement at the end produces the same result for all branches. M\"{u}ller's Theorem provides concrete limitations to this computational power. 
\begin{quote}
\textit{Interaction between branches provides no benefit in compressing classical information.}
\end{quote}


\begin{thebibliography}{Mue07}
	
	\bibitem[BV93]{BernsteinVa93}
	E.~Bernstein and U.~Vazirani.
	\newblock Quantum complexity theory.
	\newblock In {\em Proceedings of the Twenty-Fifth Annual ACM Symposium on
		Theory of Computing}, page 11–20, New York, NY, USA, 1993. Association for
	Computing Machinery.
	
	\bibitem[BvL01]{BerthiaumeVaLa01}
	A.~{Berthiaume}, W.~{van Dam}, and S.~Laplante.
	\newblock {Quantum Kolmogorov Complexity}.
	\newblock {\em Journal of Computer and System Sciences}, 63(2), 2001.
	
	\bibitem[G\'01]{Gacs01}
	P.~G\'acs.
	\newblock {Quantum Algorithmic Entropy}.
	\newblock {\em Journal of Physics A Mathematical General}, 34(35), 2001.
	
	\bibitem[Mue07]{Muller07}
	M.~Mueller.
	\newblock Quantum kolmogorov complexity and the quantum turing machine.
	\newblock {\em CoRR}, abs/0712.4377, 2007.
	
	\bibitem[Mul08]{Muller08}
	M.~Muller.
	\newblock {Strongly Universal Quantum Turing Machines and Invariance of
		Kolmogorov Complexity}.
	\newblock {\em IEEE Transactions on Information Theory}, 54(2), 2008.
	
	\bibitem[Mul09]{Muller09}
	M.~Muller.
	\newblock On the quantum kolmogorov complexity of classical strings.
	\newblock {\em International Journal of Quantum Information}, 07(04):701--711,
	2009.
	
	\bibitem[Tao]{Tao24}
	T.~Tao.
	\newblock What's new: A cheap version of the kabatjanskii-levenstein bound for
	almost orthogonal vectors.
	\newblock
	\url{https://terrytao.wordpress.com/2013/07/18/a-cheap-version-of-the-kabatjanskii-levenstein-bound-for-almost-orthogonal-vectors/}.
	\newblock Accessed: 2024-01-11.
	
\end{thebibliography}
\end{document}